\documentclass[%
 reprint,
 amsmath,amssymb,
 aip,apl,
]{revtex4-1}

\usepackage{graphicx}
\usepackage{dcolumn}
\usepackage{bm}
\usepackage[usenames,dvipsnames]{xcolor}
\usepackage{soul}
\usepackage[normalem]{ulem}

\newcommand{\pe}[1]{\textcolor{Brown}{#1}}

\begin{document}

\preprint{APS/123-QED}

\title{
Real-time sparse-sampled Ptychographic imaging through deep neural networks
}

\author{Mathew J. Cherukara}
\email{mcherukara@anl.gov}
\author{Tao Zhou}
\affiliation{%
 Center for Nanoscale Materials, Argonne National Laboratory, Lemont, IL 60439, USA\\
}%
\author{Youssef Nashed}
\affiliation{Stats Perform, Chicago, IL 60601}
\author{Pablo Enfedaque}
\affiliation{CAMERA, Lawrence Berkeley National Laboratory, Berkeley, CA}
\author{Alex Hexemer}
\affiliation{CAMERA, Lawrence Berkeley National Laboratory, Berkeley, CA}
\author{Ross J. Harder}%
\affiliation{%
 Advanced Photon Source, Argonne National Laboratory, Lemont, IL 60439, USA\\
}%
\author{Martin V. Holt}
\affiliation{%
 Center for Nanoscale Materials, Argonne National Laboratory, Lemont, IL 60439, USA\\
}%

\date{\today}

\begin{abstract}

Ptychography has rapidly grown in the fields of X-ray and electron imaging for its unprecedented ability to achieve nano or atomic scale resolution while simultaneously retrieving chemical or magnetic information from a sample. 
A ptychographic reconstruction is achieved by means of solving a complex inverse problem that imposes constraints both on the acquisition and on the analysis of the data, which typically precludes real-time imaging due to computational cost involved in solving this inverse problem.
In this work we propose PtychoNN, a novel approach to solve the ptychography reconstruction problem based on deep convolutional neural networks. We demonstrate how the proposed method can be used to predict real-space structure and phase at each scan point solely from the corresponding far-field diffraction data.
The presented results demonstrate how PtychoNN can effectively be used on experimental data, being able to generate high quality reconstructions of a sample up to hundreds of times faster than state-of-the-art ptychography reconstruction solutions once trained.
By surpassing the typical constraints of iterative model-based methods, we can significantly relax the data acquisition sampling conditions and produce equally satisfactory reconstructions. Besides drastically accelerating acquisition and analysis, this capability can enable new imaging scenarios that were not possible before, in cases of dose sensitive, dynamic and extremely voluminous samples.




\end{abstract}

\maketitle



Ptychography has emerged as a versatile imaging technique that is used with both X-ray and electron sources, in scientific fields as diverse as cell biology, materials science or electronics. 
X-ray ptychography is well developed and widely used, with multiple beamlines dedicated to it at different synchrotron sources across the world.
Comparatively, electron ptychography is performed at fewer facilities but it has lead to remarkable results on a variety of scientific scenarios, recently achieving sub-angstrom resolution\cite{Jiang2018ElectronResolution} or nanoscale 3-D imaging\cite{Gao2017ElectronImaging}.
Thanks to its ability to characterize thick samples with exceptional high resolution and requiring minimum sample preparation, among other features, ptychography has provided unprecedented insight into countless material and biological specimens. Examples include few nm imaging of integrated circuits\cite{Holler2017High-resolutionCircuits}, high resolution imaging of algae\cite{Deng2018CorrelativeAlgae} and stereocilia actin \cite{Piazza2014RevealingNanoimaging}, strain imaging of nanowires\cite{Hruszkewycz2017High-resolutionPtychography} and semiconductor heterostructures with Bragg \pe{p}tychography\cite{Holt2014StrainPtychography}.

Ptychographic imaging is performed by scanning a coherent beam across the sample while measuring the scattered intensities in the far field. Subsequently, the object is recovered by algorithmically inverting the measured coherent diffraction images. Successful inversion (or image reconstruction) of ptychographic imaging data requires the solution of a complex inverse problem, commonly referred to as phase retrieval, which consists of recovering lost phase information from measured intensities alone. Currently, the ptychography phase retrieval problem is solved using model-based iterative methods that are very computationally expensive, precluding in many cases real-time imaging\cite{Datta2019ComputationalReconstruction}. In addition, the convergence of these techniques can be extremely sensitive to the specific algorithm employed, and also on its parameters, such as the initial guess of the probe and sample, which can also perform differently depending on the characteristics and prepossessing of a beamline. Iterative model-based methods also require a large degree of oversampling to successfully converge, i.e. adjacent measured scan points need to overlap by at least 50\%.  Considering the overlapping is in 2D, this constraint can drastically limit the area or volume of the sample that can be scanned in a given amount of time. 

Neural networks have been described as universal approximators that can represent complex and abstract functions and relationships\cite{Hornik1991ApproximationNetworks}. As a result, neural networks and deep neural networks in particular have been applied successfully to a variety of problems in computer vision, natural language processing and autonomous control\cite{Lecun2015DeepLearning}. Specific to the problem of image reconstruction, deep neural networks have been used to invert magnetic resonance imaging (MRI) data\cite{Zhu2018ImageLearning}, coherent imaging data in the far-field\cite{Cherukara2018Real-timeNetworks} and holographic imaging data\cite{Rivenson2018PhaseNetworks}.

In this letter we present PtychoNN, a deep convolutional neural network that learns a direct mapping from far-field coherent diffraction data into real-space object structure and phase, and demonstrate its practical application and training on x-ray ptychographic experimental data. Our results show that, once trained, PtychoNN is up to hundreds of times faster than Ptycholib\cite{Nashed2014ParallelReconstruction}, a production-ready high performance ptychography software. In addition, since PtychoNN learns a direct relation between diffraction data and object structure and phase, overlap constraints are no longer required for successful data inversion, further accelerating reconstruction and also data acquisition by a factor of 5.

Pytchographic measurements were obtained at the X-ray nanoprobe beamline at sector 26 of the Advanced Photon Source. A tungsten test pattern etched with random features was scanned by a 60 nm coherent beam that was focused by a Fresnel zone plate. A scan of 161x161 points was acquired in steps of 30 nm which corresponds to 50\% spatial overlap. At each scan point, coherently scattered data was acquired in the far field using a Medipix3 area detector with 55$\mu m$ pixel size  at  900 cm downstream of the sample. Real-space images of amplitude and phase were subsequently recovered from the coherently scattered data using 400 iterations of ePIE as implemented in the Ptycholib package.

Figure 1 shows the structure of PtychoNN, a deep convolutional network that takes as input the raw X-ray scattering data and outputs both structure and phase. The neural network architecture consists of 3 parts, an encoder arm that learns a representation (encoding) in feature space of the input X-ray scattering data and 2 decoder arms that learn to map from the underlying manifold of the input data to real-space amplitude and phase, respectively. The encoder arm consists of convolutional and max pooling layers, and is designed to learn representations of the data at different hierarchical levels. Conversely, the decoder arms contain convolutional and upsampling layers that are designed to generate real-space amplitude and phase from the feature representation of the data provided by the encoder arm. 

\begin{figure}
\includegraphics[width=0.45\textwidth]{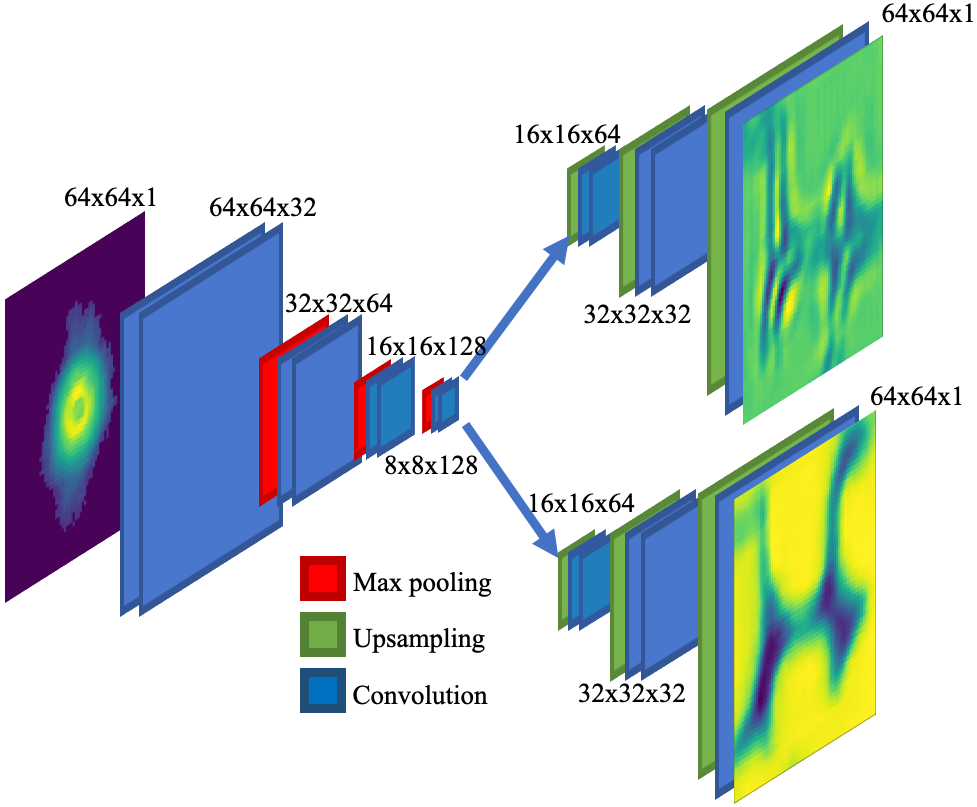}
\caption{Architecture of PtychoNN, a deep convolutional neural network that can predict real-space amplitude and phase from input diffraction data alone.}
\end{figure}

To train the network, we use the reconstruction obtained through iterative phase retrieval from the first 100 lines of the scan (see supplementary Figure 1). Hence, the training set consists of 16,100 triplets of raw X-ray scattering data, real-space amplitude and real-space phase. The training data is split 90-10 into training and validation, and the weights of the network are updated to minimize per-pixel mean absolute error (MAE). Weight updates are made using adaptive moment estimation (ADAM) with a starting learning rate of 0.001\cite{Kingma2015Adam:Optimization}. The learning rate is halved whenever training performance plateaus, i.e. when validation loss does not decrease over 5 training epochs. Each epoch represents one entire pass over the training data. Training continues for several epochs until a minimum was observed in the validation loss. Once trained, we evaluate the performance of the network on the remaining portion of the scan, i.e. on the last 61 lines of the scan (see supplementary Figure 1).

Figure 2 shows single-shot examples of the performance of PtychoNN on data from the second portion of the experimental scan, i.e. data that the network did not see during its training. The figure shows the original diffraction data, as well as the reconstructions achieved by ePIE and PtychoNN, over a 640 nm field of view dataset. The results demonstrate how  PtychoNN is able to accurately predict real-space amplitude and phase from input X-ray diffraction data alone. We also note that even though the full width half max (FWHM) of the beam is $\sim$60 nm, PtychoNN can accurately reproduce a 640 nm field of view from a single diffraction data point. This is a result of PtychoNN's ability to take advantage of information contained in the tails of the beam which extends several hundred nm (supplementary Figure 2). To recover the entire test scan, we average PtychoNN's predictions from each scan point (which are spaced 30 nm apart). Figure 3 shows PtychoNN's average predictions of amplitude and phase over the entire test area (1.8$\mu m\times$1.8$\mu m$), as well as the respective reconstruction using ePIE. The results demonstrate how PtychoNN's predictions are remarkably accurate when compared to those obtained by ePIE, while also being $\sim$300 times faster.

\begin{figure}
\includegraphics[width=0.47\textwidth]{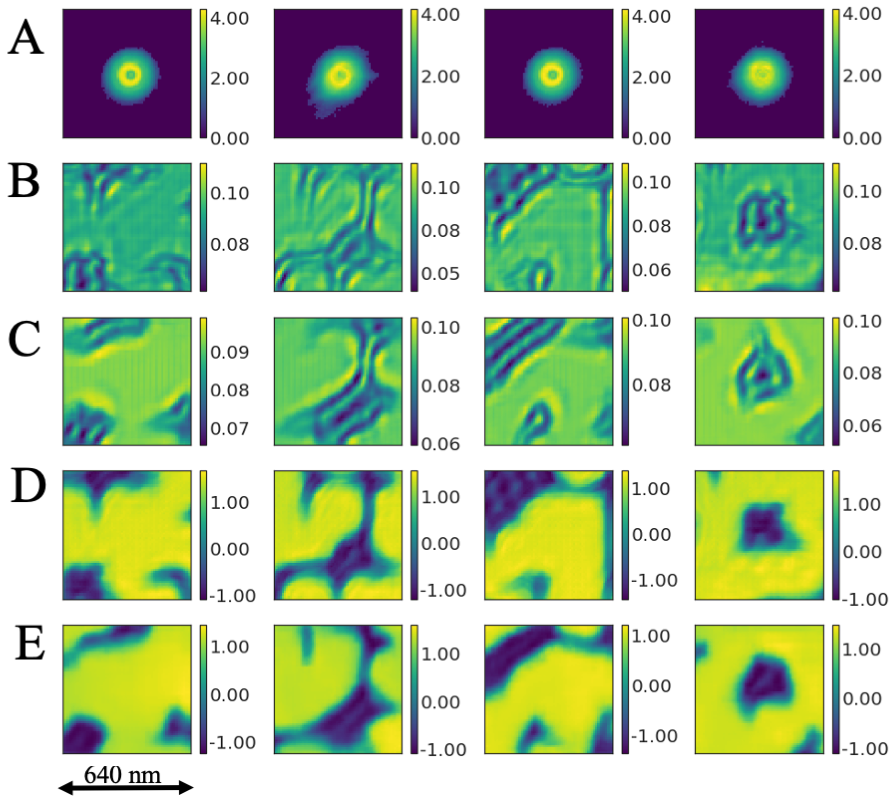}
\caption{Single-shot PtychoNN predictions on experimental test data. A) Input diffraction at different scan points, B) amplitude obtained using ePIE, C) amplitude predicted by PtychoNN, D) phase obtained using ePIE, E) phase predicted by PtychoNN.}
\end{figure}

\begin{figure}
\includegraphics[width=0.4\textwidth]{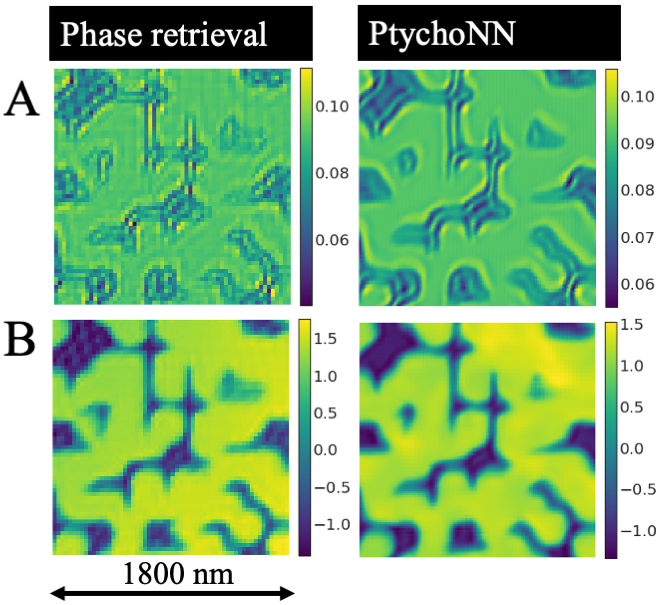}
\caption{PtychoNN performance on experimental test data. A and B show the amplitude and phase obtained from ePIE and from PtychoNN, respectively.}
\end{figure}

It is important to note that PtychoNN learns a direct mapping from reciprocal space data to real-space amplitude and phase without benefiting at all from the 50\% overlap condition, so we can now explore the possibility of PtychoNN being able to invert sparse-sampled Pytchographic data. Figure 4 shows a comparison of the performance of PtychoNN and iterative phase retrieval for different sampling conditions. The results show how, when using a 30 nm step size, the necessary overlap condition for a standard ptychographic reconstruction is satisfied and ePIE is able to successfully retrieve accurate amplitude and phase information from the diffraction data. However, if we attempt to perform iterative phase retrieval with data that has less than 50\% overlap, we begin to notice the presence of artifacts, first in the retrieved intensity (Fig. 4 A), and then also in the phase (Fig. 4 C). In contrast, the amplitude and phase predicted by PtychoNN remains remarkably accurate even when the data is sub-sampled by a factor of 5 (Fig. B and D). 

This capability of PtychoNN to successfully invert drastically sparse-sampled ptychographic data can be extremely beneficial, especially when dealing with dose-sensitive, dynamic or extremely large samples. By reducing the density of points that need to be sampled, PtychoNN can significantly reduce the radiation dose needed to image at a given resolution. Similarly, while it is always desirable to minimize data acquisition time, this is particularly vital in the case of dynamic samples in order to capture transient phenomena. We note that this sparse-sampled approach is achieved without changing the focal condition of the optic, which is currently the only means of switching between different fields of view. 

\begin{figure}
\includegraphics[width=0.47\textwidth]{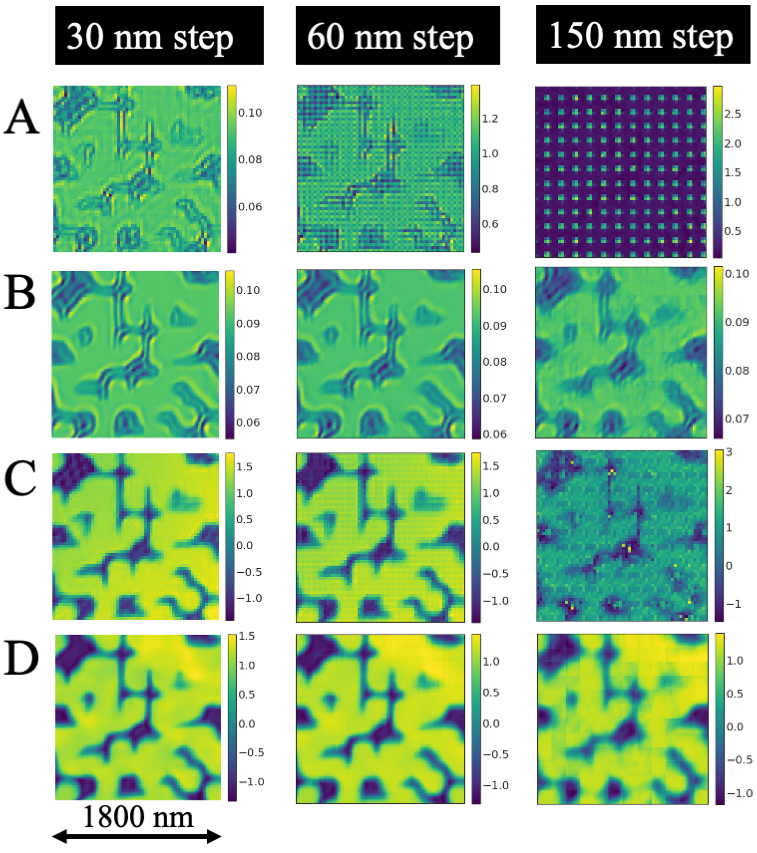}
\caption{PtychoNN performance on experimental test data under different sampling conditions. A, C show amplitude and phase obtained using ePIE. B, D show amplitude and phase predicted by PtychoNN.}
\end{figure}

Finally, we turn our attention to the question of how much training data is needed in order to obtain reasonably accurate results. Typically, training of deep convolutional neural networks often requires millions of training examples and enormous computational resources, leading to days or weeks of training\cite{Szegedy2017Inception-v4Learning}. In the results presented so far, we used 16,000 training examples to train the network; below we evaluate the performance of PtychoNN when less training data is available. Fig. 5 shows the performance of PtychoNN when trained on progressively fewer training examples, from left to right. The results show how PtychoNN can generate reasonable predictions when trained on as few as 800 training samples. Training on this small set was achieved in less than a minute on a single NVIDIA V100 GPU. The robustness of the network even when employing very reduced training sets potentially allows us to train PtychoNN on-the-fly, even on limited computational resources. Once trained, PtychoNN only takes $\sim$1 ms to make a prediction of the amplitude and phase from the diffraction data at each scan point. 

\begin{figure*}
\includegraphics[width=0.9\textwidth]{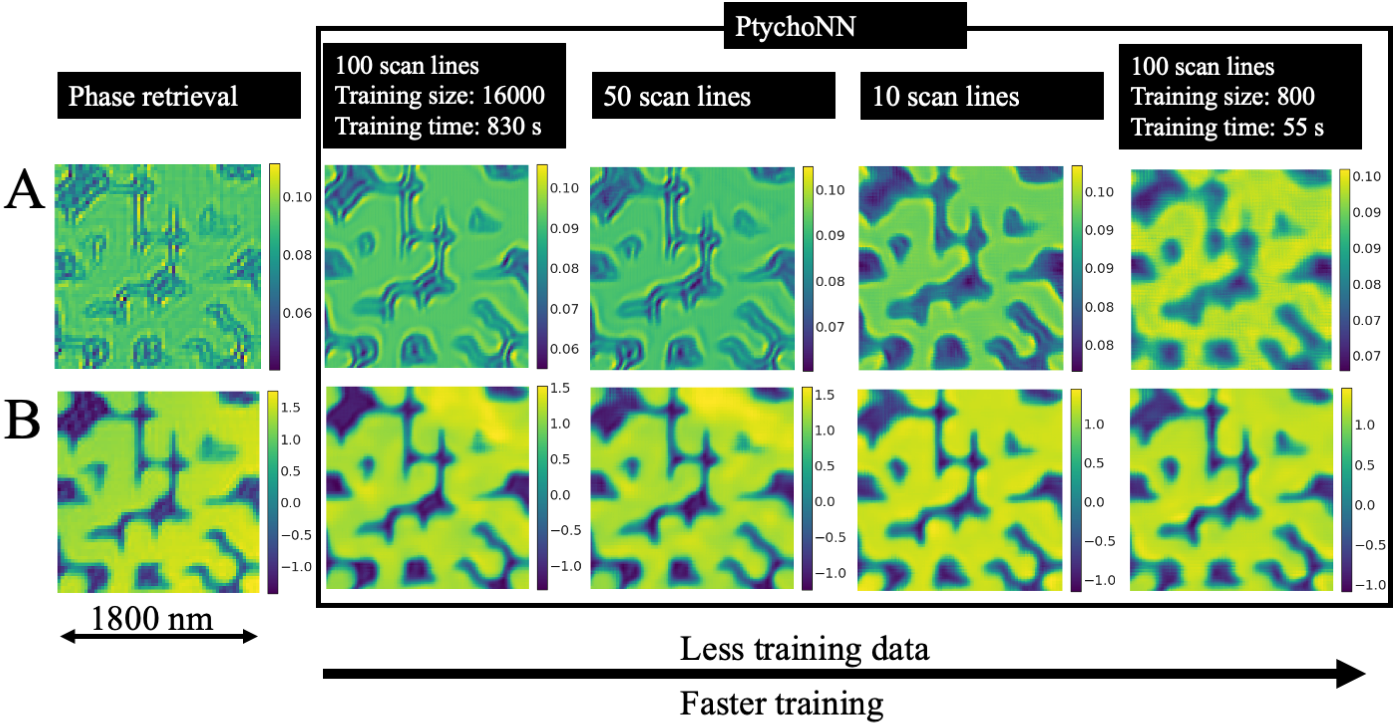}
\caption{Effect of training data size on performance. A amplitude and B phase using ePIE and PtychoNN, respectively. Images from left to right show performance of PtychoNN when trained on progressively fewer training samples.}
\end{figure*}


To the best of our knowledge, this work presents the first end-to-end machine learning solution based on convolutional deep neural networks to solve the ptychography phase retrieval problem on experimental data. We note that a similar approach has been demonstrated on simulated ptychographic data under the assumption of a known probe function\cite{Guan2019PtychoNet:Ptychography}. Our experimental results report that PtychoNN is up to 300 times faster than Ptycholib, a high performance GPU-accelerated iterative phase retrieval solution, with the additional benefit that it can be trained and deployed on minimal computing resources on the edge. In addition to remarkably faster experimental feedback, PtychoNN's ability to recover accurate real-space images from sub-sampled Ptychographic data has the potential to revolutionize ptychography dose sensitive, dynamic and extremely voluminous samples. 

We believe the results in this letter have widespread ramifications for both x-ray and electron Ptychographic imaging experiments especially in the light of increased data rates associated with faster detectors and brighter sources\cite{Tate2016HighMicroscopy,Miao2015}. 
Coherent imaging techniques, including ptychography, are the primary driver for several major upgrades to synchrotron sources across the world, including the Advanced Photon Source Upgrade (APS-U), the European Synchrotron Research Facility Extremely Brilliant Source (ESRF-EBS) and PETRA-III. Further development on deep learning methods for data inversion such as PtychoNN will be fundamental to make maximum use of these vast infrastructure upgrades and keep pace with the forthcoming computational challenges. 


This work was performed, in part, at the Center for Nanoscale Materials. Use of the Center for Nanoscale Materials and Advanced Photon Source, both Office of Science user facilities, was supported by the U.S. Department of Energy, Office of Science, Office of Basic Energy Sciences, under Contract No. DE-AC02-06CH11357. This work was also supported by Argonne LDRD 2018-019-N0:  A.I C.D.I: Atomistically Informed Coherent Diffraction Imaging. 

\bibliography{references.bib}

\begin{thebibliography}{19}%
\makeatletter
\providecommand \@ifxundefined [1]{%
 \@ifx{#1\undefined}
}%
\providecommand \@ifnum [1]{%
 \ifnum #1\expandafter \@firstoftwo
 \else \expandafter \@secondoftwo
 \fi
}%
\providecommand \@ifx [1]{%
 \ifx #1\expandafter \@firstoftwo
 \else \expandafter \@secondoftwo
 \fi
}%
\providecommand \natexlab [1]{#1}%
\providecommand \enquote  [1]{``#1''}%
\providecommand \bibnamefont  [1]{#1}%
\providecommand \bibfnamefont [1]{#1}%
\providecommand \citenamefont [1]{#1}%
\providecommand \href@noop [0]{\@secondoftwo}%
\providecommand \href [0]{\begingroup \@sanitize@url \@href}%
\providecommand \@href[1]{\@@startlink{#1}\@@href}%
\providecommand \@@href[1]{\endgroup#1\@@endlink}%
\providecommand \@sanitize@url [0]{\catcode `\\12\catcode `\$12\catcode
  `\&12\catcode `\#12\catcode `\^12\catcode `\_12\catcode `\%12\relax}%
\providecommand \@@startlink[1]{}%
\providecommand \@@endlink[0]{}%
\providecommand \url  [0]{\begingroup\@sanitize@url \@url }%
\providecommand \@url [1]{\endgroup\@href {#1}{\urlprefix }}%
\providecommand \urlprefix  [0]{URL }%
\providecommand \Eprint [0]{\href }%
\providecommand \doibase [0]{http://dx.doi.org/}%
\providecommand \selectlanguage [0]{\@gobble}%
\providecommand \bibinfo  [0]{\@secondoftwo}%
\providecommand \bibfield  [0]{\@secondoftwo}%
\providecommand \translation [1]{[#1]}%
\providecommand \BibitemOpen [0]{}%
\providecommand \bibitemStop [0]{}%
\providecommand \bibitemNoStop [0]{.\EOS\space}%
\providecommand \EOS [0]{\spacefactor3000\relax}%
\providecommand \BibitemShut  [1]{\csname bibitem#1\endcsname}%
\let\auto@bib@innerbib\@empty
\bibitem [{\citenamefont {Jiang}\ \emph {et~al.}(2018)\citenamefont {Jiang},
  \citenamefont {Chen}, \citenamefont {Han}, \citenamefont {Deb}, \citenamefont
  {Gao}, \citenamefont {Xie}, \citenamefont {Purohit}, \citenamefont {Tate},
  \citenamefont {Park}, \citenamefont {Gruner}, \citenamefont {Elser},\ and\
  \citenamefont {Muller}}]{Jiang2018ElectronResolution}%
  \BibitemOpen
  \bibfield  {author} {\bibinfo {author} {\bibfnamefont {Y.}~\bibnamefont
  {Jiang}}, \bibinfo {author} {\bibfnamefont {Z.}~\bibnamefont {Chen}},
  \bibinfo {author} {\bibfnamefont {Y.}~\bibnamefont {Han}}, \bibinfo {author}
  {\bibfnamefont {P.}~\bibnamefont {Deb}}, \bibinfo {author} {\bibfnamefont
  {H.}~\bibnamefont {Gao}}, \bibinfo {author} {\bibfnamefont {S.}~\bibnamefont
  {Xie}}, \bibinfo {author} {\bibfnamefont {P.}~\bibnamefont {Purohit}},
  \bibinfo {author} {\bibfnamefont {M.~W.}\ \bibnamefont {Tate}}, \bibinfo
  {author} {\bibfnamefont {J.}~\bibnamefont {Park}}, \bibinfo {author}
  {\bibfnamefont {S.~M.}\ \bibnamefont {Gruner}}, \bibinfo {author}
  {\bibfnamefont {V.}~\bibnamefont {Elser}}, \ and\ \bibinfo {author}
  {\bibfnamefont {D.~A.}\ \bibnamefont {Muller}},\ }\href {\doibase
  10.1038/s41586-018-0298-5} {\bibfield  {journal} {\bibinfo  {journal}
  {Nature}\ }\textbf {\bibinfo {volume} {559}},\ \bibinfo {pages} {343}
  (\bibinfo {year} {2018})}\BibitemShut {NoStop}%
\bibitem [{\citenamefont {Gao}\ \emph {et~al.}(2017)\citenamefont {Gao},
  \citenamefont {Wang}, \citenamefont {Zhang}, \citenamefont {Martinez},
  \citenamefont {Nellist}, \citenamefont {Pan},\ and\ \citenamefont
  {Kirkland}}]{Gao2017ElectronImaging}%
  \BibitemOpen
  \bibfield  {author} {\bibinfo {author} {\bibfnamefont {S.}~\bibnamefont
  {Gao}}, \bibinfo {author} {\bibfnamefont {P.}~\bibnamefont {Wang}}, \bibinfo
  {author} {\bibfnamefont {F.}~\bibnamefont {Zhang}}, \bibinfo {author}
  {\bibfnamefont {G.~T.}\ \bibnamefont {Martinez}}, \bibinfo {author}
  {\bibfnamefont {P.~D.}\ \bibnamefont {Nellist}}, \bibinfo {author}
  {\bibfnamefont {X.}~\bibnamefont {Pan}}, \ and\ \bibinfo {author}
  {\bibfnamefont {A.~I.}\ \bibnamefont {Kirkland}},\ }\href {\doibase
  10.1038/s41467-017-00150-1} {\bibfield  {journal} {\bibinfo  {journal}
  {Nature Communications}\ }\textbf {\bibinfo {volume} {8}},\ \bibinfo {pages}
  {163} (\bibinfo {year} {2017})}\BibitemShut {NoStop}%
\bibitem [{\citenamefont {Holler}\ \emph {et~al.}(2017)\citenamefont {Holler},
  \citenamefont {Guizar-Sicairos}, \citenamefont {Tsai}, \citenamefont
  {Dinapoli}, \citenamefont {M{\"{u}}ller}, \citenamefont {Bunk}, \citenamefont
  {Raabe},\ and\ \citenamefont {Aeppli}}]{Holler2017High-resolutionCircuits}%
  \BibitemOpen
  \bibfield  {author} {\bibinfo {author} {\bibfnamefont {M.}~\bibnamefont
  {Holler}}, \bibinfo {author} {\bibfnamefont {M.}~\bibnamefont
  {Guizar-Sicairos}}, \bibinfo {author} {\bibfnamefont {E.~H.}\ \bibnamefont
  {Tsai}}, \bibinfo {author} {\bibfnamefont {R.}~\bibnamefont {Dinapoli}},
  \bibinfo {author} {\bibfnamefont {E.}~\bibnamefont {M{\"{u}}ller}}, \bibinfo
  {author} {\bibfnamefont {O.}~\bibnamefont {Bunk}}, \bibinfo {author}
  {\bibfnamefont {J.}~\bibnamefont {Raabe}}, \ and\ \bibinfo {author}
  {\bibfnamefont {G.}~\bibnamefont {Aeppli}},\ }\href {\doibase
  10.1038/nature21698} {\bibfield  {journal} {\bibinfo  {journal} {Nature}\
  }\textbf {\bibinfo {volume} {543}},\ \bibinfo {pages} {402} (\bibinfo {year}
  {2017})}\BibitemShut {NoStop}%
\bibitem [{\citenamefont {Deng}\ \emph {et~al.}(2018)\citenamefont {Deng},
  \citenamefont {Lo}, \citenamefont {Gallagher-Jones}, \citenamefont {Chen},
  \citenamefont {Pryor}, \citenamefont {Jin}, \citenamefont {Hong},
  \citenamefont {Nashed}, \citenamefont {Vogt}, \citenamefont {Miao},\ and\
  \citenamefont {Jacobsen}}]{Deng2018CorrelativeAlgae}%
  \BibitemOpen
  \bibfield  {author} {\bibinfo {author} {\bibfnamefont {J.}~\bibnamefont
  {Deng}}, \bibinfo {author} {\bibfnamefont {Y.~H.}\ \bibnamefont {Lo}},
  \bibinfo {author} {\bibfnamefont {M.}~\bibnamefont {Gallagher-Jones}},
  \bibinfo {author} {\bibfnamefont {S.}~\bibnamefont {Chen}}, \bibinfo {author}
  {\bibfnamefont {A.}~\bibnamefont {Pryor}}, \bibinfo {author} {\bibfnamefont
  {Q.}~\bibnamefont {Jin}}, \bibinfo {author} {\bibfnamefont {Y.~P.}\
  \bibnamefont {Hong}}, \bibinfo {author} {\bibfnamefont {Y.~S.}\ \bibnamefont
  {Nashed}}, \bibinfo {author} {\bibfnamefont {S.}~\bibnamefont {Vogt}},
  \bibinfo {author} {\bibfnamefont {J.}~\bibnamefont {Miao}}, \ and\ \bibinfo
  {author} {\bibfnamefont {C.}~\bibnamefont {Jacobsen}},\ }\href {\doibase
  10.1126/sciadv.aau4548} {\bibfield  {journal} {\bibinfo  {journal} {Science
  Advances}\ }\textbf {\bibinfo {volume} {4}},\ \bibinfo {pages} {1} (\bibinfo
  {year} {2018})}\BibitemShut {NoStop}%
\bibitem [{\citenamefont {Piazza}\ \emph {et~al.}(2014)\citenamefont {Piazza},
  \citenamefont {Weinhausen}, \citenamefont {Diaz}, \citenamefont {Dammann},
  \citenamefont {Maurer}, \citenamefont {Reynolds}, \citenamefont
  {Burghammer},\ and\ \citenamefont
  {K{\"{o}}ster}}]{Piazza2014RevealingNanoimaging}%
  \BibitemOpen
  \bibfield  {author} {\bibinfo {author} {\bibfnamefont {V.}~\bibnamefont
  {Piazza}}, \bibinfo {author} {\bibfnamefont {B.}~\bibnamefont {Weinhausen}},
  \bibinfo {author} {\bibfnamefont {A.}~\bibnamefont {Diaz}}, \bibinfo {author}
  {\bibfnamefont {C.}~\bibnamefont {Dammann}}, \bibinfo {author} {\bibfnamefont
  {C.}~\bibnamefont {Maurer}}, \bibinfo {author} {\bibfnamefont
  {M.}~\bibnamefont {Reynolds}}, \bibinfo {author} {\bibfnamefont
  {M.}~\bibnamefont {Burghammer}}, \ and\ \bibinfo {author} {\bibfnamefont
  {S.}~\bibnamefont {K{\"{o}}ster}},\ }\href {\doibase 10.1021/nn5041526}
  {\bibfield  {journal} {\bibinfo  {journal} {ACS Nano}\ }\textbf {\bibinfo
  {volume} {8}},\ \bibinfo {pages} {12228} (\bibinfo {year}
  {2014})}\BibitemShut {NoStop}%
\bibitem [{\citenamefont {Hruszkewycz}\ \emph {et~al.}(2017)\citenamefont
  {Hruszkewycz}, \citenamefont {Allain}, \citenamefont {Holt}, \citenamefont
  {Murray}, \citenamefont {Holt}, \citenamefont {Fuoss},\ and\ \citenamefont
  {Chamard}}]{Hruszkewycz2017High-resolutionPtychography}%
  \BibitemOpen
  \bibfield  {author} {\bibinfo {author} {\bibfnamefont {S.~O.}\ \bibnamefont
  {Hruszkewycz}}, \bibinfo {author} {\bibfnamefont {M.}~\bibnamefont {Allain}},
  \bibinfo {author} {\bibfnamefont {M.~V.}\ \bibnamefont {Holt}}, \bibinfo
  {author} {\bibfnamefont {C.~E.}\ \bibnamefont {Murray}}, \bibinfo {author}
  {\bibfnamefont {J.~R.}\ \bibnamefont {Holt}}, \bibinfo {author}
  {\bibfnamefont {P.~H.}\ \bibnamefont {Fuoss}}, \ and\ \bibinfo {author}
  {\bibfnamefont {V.}~\bibnamefont {Chamard}},\ }\href {\doibase
  10.1038/nmat4798} {\bibfield  {journal} {\bibinfo  {journal} {Nature
  Materials}\ }\textbf {\bibinfo {volume} {16}},\ \bibinfo {pages} {244}
  (\bibinfo {year} {2017})}\BibitemShut {NoStop}%
\bibitem [{\citenamefont {Holt}\ \emph {et~al.}(2014)\citenamefont {Holt},
  \citenamefont {Hruszkewycz}, \citenamefont {Murray}, \citenamefont {Holt},
  \citenamefont {Paskiewicz},\ and\ \citenamefont
  {Fuoss}}]{Holt2014StrainPtychography}%
  \BibitemOpen
  \bibfield  {author} {\bibinfo {author} {\bibfnamefont {M.~V.}\ \bibnamefont
  {Holt}}, \bibinfo {author} {\bibfnamefont {S.~O.}\ \bibnamefont
  {Hruszkewycz}}, \bibinfo {author} {\bibfnamefont {C.~E.}\ \bibnamefont
  {Murray}}, \bibinfo {author} {\bibfnamefont {J.~R.}\ \bibnamefont {Holt}},
  \bibinfo {author} {\bibfnamefont {D.~M.}\ \bibnamefont {Paskiewicz}}, \ and\
  \bibinfo {author} {\bibfnamefont {P.~H.}\ \bibnamefont {Fuoss}},\ }\href
  {\doibase 10.1103/PhysRevLett.112.165502} {\bibfield  {journal} {\bibinfo
  {journal} {Physical Review Letters}\ }\textbf {\bibinfo {volume} {112}},\
  \bibinfo {pages} {1} (\bibinfo {year} {2014})}\BibitemShut {NoStop}%
\bibitem [{\citenamefont {Datta}\ \emph {et~al.}(2019)\citenamefont {Datta},
  \citenamefont {Rittenbach}, \citenamefont {Kang}, \citenamefont {Walters},
  \citenamefont {Crago},\ and\ \citenamefont
  {Damoulakis}}]{Datta2019ComputationalReconstruction}%
  \BibitemOpen
  \bibfield  {author} {\bibinfo {author} {\bibfnamefont {K.}~\bibnamefont
  {Datta}}, \bibinfo {author} {\bibfnamefont {A.}~\bibnamefont {Rittenbach}},
  \bibinfo {author} {\bibfnamefont {D.-I.}\ \bibnamefont {Kang}}, \bibinfo
  {author} {\bibfnamefont {J.~P.}\ \bibnamefont {Walters}}, \bibinfo {author}
  {\bibfnamefont {S.~P.}\ \bibnamefont {Crago}}, \ and\ \bibinfo {author}
  {\bibfnamefont {J.}~\bibnamefont {Damoulakis}},\ }\href {\doibase
  10.1364/ao.58.000b19} {\bibfield  {journal} {\bibinfo  {journal} {Applied
  Optics}\ }\textbf {\bibinfo {volume} {58}},\ \bibinfo {pages} {B19} (\bibinfo
  {year} {2019})}\BibitemShut {NoStop}%
\bibitem [{\citenamefont {Hornik}(1991)}]{Hornik1991ApproximationNetworks}%
  \BibitemOpen
  \bibfield  {author} {\bibinfo {author} {\bibfnamefont {K.}~\bibnamefont
  {Hornik}},\ }\href {\doibase 10.1016/0893-6080(91)90009-T} {\bibfield
  {journal} {\bibinfo  {journal} {Neural Networks}\ }\textbf {\bibinfo {volume}
  {4}},\ \bibinfo {pages} {251} (\bibinfo {year} {1991})}\BibitemShut {NoStop}%
\bibitem [{\citenamefont {Lecun}, \citenamefont {Bengio},\ and\ \citenamefont
  {Hinton}(2015)}]{Lecun2015DeepLearning}%
  \BibitemOpen
  \bibfield  {author} {\bibinfo {author} {\bibfnamefont {Y.}~\bibnamefont
  {Lecun}}, \bibinfo {author} {\bibfnamefont {Y.}~\bibnamefont {Bengio}}, \
  and\ \bibinfo {author} {\bibfnamefont {G.}~\bibnamefont {Hinton}},\ }\href
  {\doibase 10.1038/nature14539} {\bibfield  {journal} {\bibinfo  {journal}
  {Nature}\ }\textbf {\bibinfo {volume} {521}},\ \bibinfo {pages} {436}
  (\bibinfo {year} {2015})}\BibitemShut {NoStop}%
\bibitem [{\citenamefont {Zhu}\ \emph {et~al.}(2018)\citenamefont {Zhu},
  \citenamefont {Liu}, \citenamefont {Cauley}, \citenamefont {Rosen},\ and\
  \citenamefont {Rosen}}]{Zhu2018ImageLearning}%
  \BibitemOpen
  \bibfield  {author} {\bibinfo {author} {\bibfnamefont {B.}~\bibnamefont
  {Zhu}}, \bibinfo {author} {\bibfnamefont {J.~Z.}\ \bibnamefont {Liu}},
  \bibinfo {author} {\bibfnamefont {S.~F.}\ \bibnamefont {Cauley}}, \bibinfo
  {author} {\bibfnamefont {B.~R.}\ \bibnamefont {Rosen}}, \ and\ \bibinfo
  {author} {\bibfnamefont {M.~S.}\ \bibnamefont {Rosen}},\ }\href {\doibase
  10.1038/nature25988} {\bibfield  {journal} {\bibinfo  {journal} {Nature}\
  }\textbf {\bibinfo {volume} {555}},\ \bibinfo {pages} {487} (\bibinfo {year}
  {2018})}\BibitemShut {NoStop}%
\bibitem [{\citenamefont {Cherukara}, \citenamefont {Nashed},\ and\
  \citenamefont {Harder}(2018)}]{Cherukara2018Real-timeNetworks}%
  \BibitemOpen
  \bibfield  {author} {\bibinfo {author} {\bibfnamefont {M.~J.}\ \bibnamefont
  {Cherukara}}, \bibinfo {author} {\bibfnamefont {Y.~S.}\ \bibnamefont
  {Nashed}}, \ and\ \bibinfo {author} {\bibfnamefont {R.~J.}\ \bibnamefont
  {Harder}},\ }\href {\doibase 10.1038/s41598-018-34525-1} {\bibfield
  {journal} {\bibinfo  {journal} {Scientific Reports}\ }\textbf {\bibinfo
  {volume} {8}},\ \bibinfo {pages} {1} (\bibinfo {year} {2018})}\BibitemShut
  {NoStop}%
\bibitem [{\citenamefont {Rivenson}\ \emph {et~al.}(2018)\citenamefont
  {Rivenson}, \citenamefont {Zhang}, \citenamefont {G{\"{u}}naydın},
  \citenamefont {Teng},\ and\ \citenamefont
  {Ozcan}}]{Rivenson2018PhaseNetworks}%
  \BibitemOpen
  \bibfield  {author} {\bibinfo {author} {\bibfnamefont {Y.}~\bibnamefont
  {Rivenson}}, \bibinfo {author} {\bibfnamefont {Y.}~\bibnamefont {Zhang}},
  \bibinfo {author} {\bibfnamefont {H.}~\bibnamefont {G{\"{u}}naydın}},
  \bibinfo {author} {\bibfnamefont {D.}~\bibnamefont {Teng}}, \ and\ \bibinfo
  {author} {\bibfnamefont {A.}~\bibnamefont {Ozcan}},\ }\href {\doibase
  10.1038/lsa.2017.141} {\bibfield  {journal} {\bibinfo  {journal} {Light:
  Science {\&} Applications}\ }\textbf {\bibinfo {volume} {7}},\ \bibinfo
  {pages} {17141} (\bibinfo {year} {2018})}\BibitemShut {NoStop}%
\bibitem [{\citenamefont {Nashed}\ \emph {et~al.}(2014)\citenamefont {Nashed},
  \citenamefont {Vine}, \citenamefont {Peterka}, \citenamefont {Deng},
  \citenamefont {Ross},\ and\ \citenamefont
  {Jacobsen}}]{Nashed2014ParallelReconstruction}%
  \BibitemOpen
  \bibfield  {author} {\bibinfo {author} {\bibfnamefont {Y.~S.~G.}\
  \bibnamefont {Nashed}}, \bibinfo {author} {\bibfnamefont {D.~J.}\
  \bibnamefont {Vine}}, \bibinfo {author} {\bibfnamefont {T.}~\bibnamefont
  {Peterka}}, \bibinfo {author} {\bibfnamefont {J.}~\bibnamefont {Deng}},
  \bibinfo {author} {\bibfnamefont {R.}~\bibnamefont {Ross}}, \ and\ \bibinfo
  {author} {\bibfnamefont {C.}~\bibnamefont {Jacobsen}},\ }\href {\doibase
  10.1364/oe.22.032082} {\bibfield  {journal} {\bibinfo  {journal} {Optics
  Express}\ }\textbf {\bibinfo {volume} {22}},\ \bibinfo {pages} {32082}
  (\bibinfo {year} {2014})}\BibitemShut {NoStop}%
\bibitem [{\citenamefont {Kingma}\ and\ \citenamefont
  {Ba}(2015)}]{Kingma2015Adam:Optimization}%
  \BibitemOpen
  \bibfield  {author} {\bibinfo {author} {\bibfnamefont {D.~P.}\ \bibnamefont
  {Kingma}}\ and\ \bibinfo {author} {\bibfnamefont {J.}~\bibnamefont {Ba}},\
  }in\ \href {http://arxiv.org/abs/1412.6980
  http://portal.acm.org/citation.cfm?doid=1830483.1830503} {\emph {\bibinfo
  {booktitle} {Proceedings of the 3rd International Conference on Learning
  Representations (ICLR)}}}\ (\bibinfo {year} {2015})\BibitemShut {NoStop}%
\bibitem [{\citenamefont {Szegedy}\ \emph {et~al.}(2017)\citenamefont
  {Szegedy}, \citenamefont {Ioffe}, \citenamefont {Vanhoucke},\ and\
  \citenamefont {Alemi}}]{Szegedy2017Inception-v4Learning}%
  \BibitemOpen
  \bibfield  {author} {\bibinfo {author} {\bibfnamefont {C.}~\bibnamefont
  {Szegedy}}, \bibinfo {author} {\bibfnamefont {S.}~\bibnamefont {Ioffe}},
  \bibinfo {author} {\bibfnamefont {V.}~\bibnamefont {Vanhoucke}}, \ and\
  \bibinfo {author} {\bibfnamefont {A.}~\bibnamefont {Alemi}},\ }in\ \href
  {http://arxiv.org/abs/1602.07261} {\emph {\bibinfo {booktitle} {Proceedings
  of the Thirty-First AAAI Conference on Artificial Intelligence (AAAI-17)}}}\
  (\bibinfo {year} {2017})\BibitemShut {NoStop}%
\bibitem [{\citenamefont {Guan}\ \emph {et~al.}(2019)\citenamefont {Guan},
  \citenamefont {Tsai}, \citenamefont {Huang}, \citenamefont {Yager},\ and\
  \citenamefont {Qin}}]{Guan2019PtychoNet:Ptychography}%
  \BibitemOpen
  \bibfield  {author} {\bibinfo {author} {\bibfnamefont {Z.}~\bibnamefont
  {Guan}}, \bibinfo {author} {\bibfnamefont {E.~H.~R.}\ \bibnamefont {Tsai}},
  \bibinfo {author} {\bibfnamefont {X.}~\bibnamefont {Huang}}, \bibinfo
  {author} {\bibfnamefont {K.~G.}\ \bibnamefont {Yager}}, \ and\ \bibinfo
  {author} {\bibfnamefont {H.}~\bibnamefont {Qin}},\ }in\ \href@noop {} {\emph
  {\bibinfo {booktitle} {British Machine Vision Conference}}}\ (\bibinfo {year}
  {2019})\ p.\ \bibinfo {pages} {1172}\BibitemShut {NoStop}%
\bibitem [{\citenamefont {Tate}\ \emph {et~al.}(2016)\citenamefont {Tate},
  \citenamefont {Purohit}, \citenamefont {Chamberlain}, \citenamefont {Nguyen},
  \citenamefont {Hovden}, \citenamefont {Chang}, \citenamefont {Deb},
  \citenamefont {Turgut}, \citenamefont {Heron}, \citenamefont {Schlom},
  \citenamefont {Ralph}, \citenamefont {Fuchs}, \citenamefont {Shanks},
  \citenamefont {Philipp}, \citenamefont {Muller},\ and\ \citenamefont
  {Gruner}}]{Tate2016HighMicroscopy}%
  \BibitemOpen
  \bibfield  {author} {\bibinfo {author} {\bibfnamefont {M.~W.}\ \bibnamefont
  {Tate}}, \bibinfo {author} {\bibfnamefont {P.}~\bibnamefont {Purohit}},
  \bibinfo {author} {\bibfnamefont {D.}~\bibnamefont {Chamberlain}}, \bibinfo
  {author} {\bibfnamefont {K.~X.}\ \bibnamefont {Nguyen}}, \bibinfo {author}
  {\bibfnamefont {R.}~\bibnamefont {Hovden}}, \bibinfo {author} {\bibfnamefont
  {C.~S.}\ \bibnamefont {Chang}}, \bibinfo {author} {\bibfnamefont
  {P.}~\bibnamefont {Deb}}, \bibinfo {author} {\bibfnamefont {E.}~\bibnamefont
  {Turgut}}, \bibinfo {author} {\bibfnamefont {J.~T.}\ \bibnamefont {Heron}},
  \bibinfo {author} {\bibfnamefont {D.~G.}\ \bibnamefont {Schlom}}, \bibinfo
  {author} {\bibfnamefont {D.~C.}\ \bibnamefont {Ralph}}, \bibinfo {author}
  {\bibfnamefont {G.~D.}\ \bibnamefont {Fuchs}}, \bibinfo {author}
  {\bibfnamefont {K.~S.}\ \bibnamefont {Shanks}}, \bibinfo {author}
  {\bibfnamefont {H.~T.}\ \bibnamefont {Philipp}}, \bibinfo {author}
  {\bibfnamefont {D.~A.}\ \bibnamefont {Muller}}, \ and\ \bibinfo {author}
  {\bibfnamefont {S.~M.}\ \bibnamefont {Gruner}},\ }\href {\doibase
  10.1017/S1431927615015664} {\bibfield  {journal} {\bibinfo  {journal}
  {Microscopy and Microanalysis}\ }\textbf {\bibinfo {volume} {22}},\ \bibinfo
  {pages} {237} (\bibinfo {year} {2016})}\BibitemShut {NoStop}%
\bibitem [{\citenamefont {Miao}\ \emph {et~al.}(2015)\citenamefont {Miao},
  \citenamefont {Ishikawa}, \citenamefont {Robinson},\ and\ \citenamefont
  {Murnane}}]{Miao2015}%
  \BibitemOpen
  \bibfield  {author} {\bibinfo {author} {\bibfnamefont {J.}~\bibnamefont
  {Miao}}, \bibinfo {author} {\bibfnamefont {T.}~\bibnamefont {Ishikawa}},
  \bibinfo {author} {\bibfnamefont {I.~K.}\ \bibnamefont {Robinson}}, \ and\
  \bibinfo {author} {\bibfnamefont {M.~M.}\ \bibnamefont {Murnane}},\ }\href
  {\doibase 10.1126/science.aaa1394} {\bibfield  {journal} {\bibinfo  {journal}
  {Science}\ }\textbf {\bibinfo {volume} {348}},\ \bibinfo {pages} {530}
  (\bibinfo {year} {2015})}\BibitemShut {NoStop}%
\end{thebibliography}%

\end{document}